\documentstyle[aps,preprint]{revtex} %
\begin{document} %
\draft %
\widetext %
\title{Collective excitations of trapped binary mixtures  %
of Bose-condensed gases} %
\author{Robert Graham\cite{permad} %
and Dan Walls} %
\address{Department of Physics, University of Auckland, \protect \\ %
Private Bag 92019, Auckland, New Zealand} %
\date{\today} %
\maketitle %
\begin{abstract} %
The linearised time dependent coupled  %
Gross-Pitaevskii equations describing  %
the long wavelength excitations of Bose-
condensed binary mixtures are solved,  %
in the bulk and in harmonic traps for the case where only the binary phase %
is present. %
In the former case we obtain two zero-sound branches  %
In the latter case   %
the dispersion law also contains two branches  %
whose dependence on the quantum numbers  %
of the modes is the same as for %
a one-component condensate,but with different prefactors, %
depending on the ratios of the three s-wave scattering lengths  %
of the two atomic species. In the general case where the binary phase %
in the trap coexists with one or both one-component phases, the  %
mode spectrum depends on the geometry of the interphase boundaries due to %
the boundary conditions there. %
Measurements of the  %
oscillation frequencies as in recent experiments with  %
modulated traps would yield very detailed  %
information on this system.  %
\end{abstract} %
The recent dramatic advances in the  %
ability to create Bose condensates have now  %
reached the stage where Bose condensates  %
of binary mixtures of two atomic species only  %
distinguished in their angular momentum  %
quantum numbers can be manufactured  %
in a magnetic trap at effectively zero temperature  %
\cite{1}. The ground state of such binary  %
Bose-condensed mixtures have recently been studied theoretically  %
in the Thomas-Fermi approximation \cite{2}. %
To learn more about their properties it  %
seems interesting  to study their long  %
wavelength excitations. Recent experiments  %
for one-component condensates  %
have been reported in \cite{3,4}, where  %
collective excitations have been excited by  %
changes in the trap potential. Theoretical  %
work on collective excitations, again for a  %
single-component condensate, has been  %
presented in \cite{5}, where the linearised  %
Gross-Pitaevskii equation was solved numerically, %
and in \cite{6}, where an analytical solution was obtained %
in the long wavelength limit. %
Very good  %
agreement between the theoretical predictions  %
for the low lying mode frequencies \cite{5,6}  %
and the experimental results \cite{3,4} was  %
obtained. As  %
shown by Stringari \cite{6}, in the Thomas Fermi
approximation, the frequencies of the  %
long wavelength excitations of a  %
trapped Bose-Einstein condensate are  %
independent of the microscopic  %
properties of the condensate, in particular of  %
the two-body scattering length, and only  %
depend on the frequency $\omega_0$ of the single-particle  %
oscillation in the trap.  %

In the present paper we shall consider  %
the two coupled linearised Gross-Pitaevskii  %
equations at zero temperature describing the excitations of  %
the binary phase of a mixture of  %
two Bose condensates. Considering the spatially homogeneous system first %
we obtain  %
the Bogoliubov spectrum for this case with  %
two branches of zero sound for $k \rightarrow 0$, where  %
the sound velocities depend on the ratios of the  %
three different scattering lengths in the problem.  %
We also derive the coupled linearised  %
hydrodynamic equations and their boundary conditions which govern  %
the long wavelength excitations of Bose %
condensed binary mixtures in a trap coexisting with the one-component %
condensates. For the special case of a pure binary-phase condensate  %
in a spherically symmetric trap we determine the mode-spectrum analytically  %
in the long-wavelength and Thomas-Fermi approximations. %
We find that the dependence of the  %
mode spectrum on the radial and  %
angular momentum quantum numbers  %
$n$ and $l$ is the same as in the case of  %
a single-component mixture,  %
\begin{eqnarray} 
\label{0}%
\omega_{+,-} = \omega_{b+,-}\left (  %
2 n^2 + 2 n l + 3 n + l %
\right )^{1/2} , %
\end{eqnarray}
but the prefactor $\omega ^2_{b+,-}$ is no longer simply  %
given by the trap frequency, $\omega_{0}$, but  %
it has two solutions, corresponding to the two  %
sound branches in the bulk, each of which  %
depends on the two ratios of the two-particle scattering lengths.  %

The coupled Gross-Pitaevskii equations %
describing a binary mixture of Bose-Einstein  %
condensates with equal particle masses are given by   %
\begin{mathletters} %
\label{1}%
\begin{eqnarray} 
i\hbar \dot \psi _1 = - \frac{\hbar ^2}{2m} %
\mbox{\boldmath $%
\nabla %
$}%
^2 \psi _1 + \psi _1 \left (  %
V_1 |\psi _1|^2 + V_{12} |\psi _2|^2  %
\right ) + U_1(%
\mbox{\boldmath $x$}%
) \psi _1 , \\ %
i\hbar \dot \psi _2 = - \frac{\hbar ^2}{2m} %
\mbox{\boldmath $%
\nabla %
$}%
^2 \psi _2 + \psi _2 \left (  %
V_2 |\psi _2|^2 + V_{12} |\psi _1|^2  %
\right ) + U_2(%
\mbox{\boldmath $x$}%
) \psi _2 . %
\end{eqnarray}
\end{mathletters}
We shall assume in this paper that the conditions  %
\begin{eqnarray} 
V_1>0, V_2>0, V_1V_2 - V_{12}^2>0  %
\end{eqnarray}
are satisfied. Here $V_i = \frac{4 \pi \hbar ^2}{m}a_i$, with  %
$i = 1, 2, 12$, are the interaction parameters, $a_i$ being the $s$-wave  %
scattering lengths. Let us first consider a spatially homogeneous  %
system, i.e. the case of vanishing trap potentials %
$U_1(%
\mbox{\boldmath $x$}%
)=U_2(%
\mbox{\boldmath $x$}%
)=0$, where  %
the solutions of (\ref{1}) of the form  %
$\psi _i \sim \text{e}^{- i \mu _i t / \hbar }$ %
are given by   %
\begin{mathletters} %
\label{2}%
\begin{eqnarray} 
\left |  %
\psi _1 %
\right | ^2 & = &  %
\left |  %
\psi _1^{(12)} %
\right | ^2 =  %
\frac{V_2 \mu _1 - V_{12}\mu _2}{V_1 V_2 - V_{12}^2} , \ \  %
\left |  %
\psi _2 %
\right | ^2  =  %
\left |  %
\psi _2^{(12)} %
\right | ^2 =  %
\frac{V_1 \mu _2 - V_{12}\mu _1}{V_1 V_2 - V_{12}^2} , \\ %
\left |  %
\psi _1 %
\right | ^2 & = &  %
\left |  %
\psi _1^{(1)} %
\right | ^2 =  %
\frac{\mu _1 }{V_1 } , \ \  %
\left |  %
\psi _2 %
\right | ^2  =  %
\left |  %
\psi _2^{(1)} %
\right | ^2 =  %
0 , \\ %
\left |  %
\psi _1 %
\right | ^2 & = &  %
\left |  %
\psi _1^{(2)} %
\right | ^2 =  %
0 , \ \  %
\left |  %
\psi _2 %
\right | ^2  =  %
\left |  %
\psi _2^{(2)} %
\right | ^2 =  %
\frac{\mu _2 }{V_2 } . %
\end{eqnarray}
\end{mathletters}

The $\mu _i$, with $i = 1,2$, are the chemical potentials  %
of the two particle species to be determined by normalising $|\psi _i|^2$  %
to the particle numbers $N_i$. The stability regions of the three solutions have been discussed in \cite{2}.  %
In particular, the solution (2) for the binary phase  %
is stable if its right-hand sides are both positive.  %

Putting now   %
\begin{eqnarray} 
\label{3}%
\psi _i = \sqrt{|\psi _i^{(12)}|^2 + \rho _i} %
\text{e}^{-i\left (  %
\mu _it/\hbar  - \varphi _i %
\right )} ,  %
\end{eqnarray}
and linearizing Eqs. (\ref{1}) in $\rho _i$  %
and $\varphi _i$ we obtain for the excitations  %
in the stability region of the binary condensate,   %
\begin{mathletters} %
\label{4}%
\begin{eqnarray} 
\dot \varphi _1 &=& -\frac{1}{\hbar }\left (  %
V_1 \rho _1 + V_{12} \rho _2  %
\right ) %
+ \frac{\hbar }{4 m |\psi _1^{(12)}|^2} %
\mbox{\boldmath $%
\nabla %
$}%
^2 \rho _1, \\ %
\dot \rho _1 &=& -\frac{\hbar }{m}|\psi _1^{(12)}|^2 %
\mbox{\boldmath $%
\nabla %
$}%
^2 \varphi _1,  %
\end{eqnarray}
\end{mathletters}
and similarly for $\varphi _2, \rho _2$ simply by interchanging  %
indices. The dispersion law for solutions $\varphi _i, \rho _i$  %
proportional to $\text{e} ^{i(\omega  t  - %
\mbox{\boldmath $k$}%
\cdot %
\mbox{\boldmath $x$}%
)}$  %
is easily obtained as  %
$\omega ^2 = \lambda _{+,-}^2(k^2)k^2$ with   %
\begin{mathletters} %
\label{5}%
\begin{eqnarray} 
\lambda _{+,-}^2(k^2) &=&  %
\frac{\hbar ^2 k^2}{(2m)^2} + c_{+,-}^2, \\ %
c_{+,-}^2 &=& \frac{2\pi \hbar ^2}{m^2}\left \{  %
a_1 |\psi _1^{(12)}|^2 + %
a_2 |\psi _2^{(12)}|^2 \pm \left [  %
\left (  %
a_1 |\psi _1^{(12)}|^2 - %
a_2 |\psi _2^{(12)}|^2 %
\right )^2 + 4 a_{12}^2 %
|\psi _1^{(12)}\psi _2^{(12)}|^2 %
\right ]^{1/2} %
\right \} %
\end{eqnarray}
\end{mathletters}
For large $k$ the two branches of the dispersion law  %
approach that for the free particles like  %
\begin{eqnarray} 
\omega &=&  %
\frac{\hbar k^2}{2m}+\frac{mc_{+,-}^2}{\hbar}  %
\end{eqnarray}
The constant shift from the free-particle dispersion law  %
would show up e.g. in light scattering as discussed for the case  %
of a single component condensate in \cite{GW} .  %
For $k \rightarrow 0$, on the other hand,  %
we obtain two zero-sound branches of collective  %
excitations with sound velocities $c_{+,-}$ %
where the densities of the two components oscillate  %
out of phase and in phase respectively.  %

Let us now turn to the case of a trapped  %
binary condensate, $U_1(%
\mbox{\boldmath $x$}%
), U_2(%
\mbox{\boldmath $x$}%
) \neq 0$, which we consider in  %
the Thomas-Fermi approximation.  %
Then the time-dependent solutions (\ref{2})  %
still apply with the replacements  %
$\mu _i \rightarrow \mu _i - U_i(%
\mbox{\boldmath $x$}%
)$(see \cite{2}). Linearising Eqs. (\ref{1})  %
with the ansatz (\ref{3}) and making a long-wavelength  %
approximation by keeping only the lowest order  %
spatial derivatives we obtain in the place of (\ref{4})   %
\begin{mathletters} %
\label{6}%
\begin{eqnarray} 
\dot \varphi _1 &=& -\frac{1}{\hbar }\left (  %
V_1 \rho _1 + V_{12} \rho _2  %
\right ) , \\ %
\dot \rho _1 &=& -\frac{\hbar }{m}%
\mbox{\boldmath $%
\nabla %
$}%
\cdot|\psi _1^{(12)}(%
\mbox{\boldmath $x$}%
)|^2 %
\mbox{\boldmath $%
\nabla %
$}%
 \varphi _1,  %
\end{eqnarray}
\end{mathletters}
and similarly for $\varphi _2, \rho _2$ by interchanging indices.  %
Eliminating the phases $\varphi _i$ we obtain  %
the two coupled equations for the long-wavelength  %
density fluctuations,   %
\begin{mathletters} %
\label{7}%
\begin{eqnarray} 
\ddot \rho _1 = \frac{1}{m} %
\mbox{\boldmath $%
\nabla %
$}%
\cdot|\psi _1^{(12)}(%
\mbox{\boldmath $x$}%
)|^2 %
\mbox{\boldmath $%
\nabla %
$}%
(V_1 \rho _1 + V_{12}\rho _2 ), \\ %
\ddot \rho _2 = \frac{1}{m} %
\mbox{\boldmath $%
\nabla %
$}%
\cdot|\psi _2^{(12)}(%
\mbox{\boldmath $x$}%
)|^2 %
\mbox{\boldmath $%
\nabla %
$}%
(V_2 \rho _2 + V_{12}\rho _1 ).  %
\end{eqnarray}
\end{mathletters}
In the general case, where the [12]-phase coexists with the one-component phases %
$k=[1],[2]$ Eqs. (\ref{7}) have to be solved simultaneously with the equations %
\begin{eqnarray} 
\ddot \rho _i = \frac{1}{m} %
\mbox{\boldmath $%
\nabla %
$}%
\cdot|\psi _i^{(k)}(%
\mbox{\boldmath $x$}%
)|^2 %
\mbox{\boldmath $%
\nabla %
$}%
V_i \rho _i   %
\end{eqnarray}
and boundary conditions. These are dictated by number conservation and %
require the continuity of $\rho_i$ and the normal derivative of $\varphi_i$ %
across the interphase boundary between the phases [12] and k on which %
$\psi_i^{(k)}\ne 0$. In general, the spectrum of collective excitations of %
the total system will therefore depend on the details of the geometry %
of the interphase %
boundaries, and solutions of the coupled equations can only be obtained %
numerically. %
Here we shall consider a special case where no interphase %
boundaries are present, for which an analytical solution can be given similar %
to the case of a one-component condensate considered in ref.\cite{6}: %
Let us consider the case where %
$U_i = \frac{1}{2}m\Omega _i^2 r^2$ represent two  concentric isotropic harmonic %
traps. It follows from the results in \cite{2} that in the special case %
where $\mu_2/\mu_1=(\Omega_2/\Omega_1)^2$ with %
\begin{mathletters} %
\label{A2}%
\begin{eqnarray} 
max(\frac{V_2}{V_{12}},\frac{V_{12}}{V_1})>\frac{\Omega_2^2}{\Omega_1^2}>min(\frac{V_2}{V_{12}},\frac{V_{12}}{V_1})
\end{eqnarray}
\end{mathletters} %
no inter-phase boundaries occur in the trap and only the binary phase is %
present there. %
The condition for the chemical potentials requires that the particle numbers %
are chosen in the ratio $N_2/N_1=(V_1\Omega_2^2-V_{12}\Omega_1^2)/(V_2\Omega_1^2-V_{12}\Omega_2^2)$, %
which can be satisfied mathematically if (\ref{A2}) holds, but can be satisfied %
physically only if the ratios of the three scattering lengths is known. %
Under these conditions we can solve  %
Eqs. (\ref{7}) by an ansatz  %
$\rho _i = \text{e}^{i\omega t} %
\sum_{k=0}^{n} C_{ik}r^{l + 2 k}Y_{lm}(\theta ,\phi )$. %
The coefficients $C_{ik}$, $i=1,2$, $k = 0, 1, \cdots, n$,  %
are determined by a recursion law. The condition $C_{i\,n+1}=0$ %
determines $\omega ^2$ by Eq. (\ref{0}) with   %
\begin{eqnarray} 
\label{8}%
\lefteqn{\omega ^2_{b\, +,-} = \frac{1}{2\left (  %
a_1 a_2 - a^2_{12} %
\right )}\left \{  %
\Omega _1^2 a_2 (a_1 - a_{12}) +  %
\Omega _2^2 a_1 (a_2 - a_{12})  %
\right . }\\ &&\left. %
\pm \left [  %
\left (  %
\Omega _1^2 a_2 (a_1 + a_{12}) -  %
\Omega _2^2 a_1 (a_2 + a_{12})   %
\right )^2 +  %
4 a_{12}^2 %
\left (  %
a_2 \Omega _1^2 - a_{12}\Omega ^2_2 %
\right ) %
\left (  %
a_1 \Omega _2^2 - a_{12}\Omega ^2_1 %
\right ) %
\right ] ^{1/2} %
\right \} .\nonumber %
\end{eqnarray}
For $a_{12} = 0$ the dispersion law (\ref{8}) reduces to Stringari's result  %
$\omega ^2_{b\, +,-} = \Omega ^2_{1,2}$ for each particle species \cite{6}.  %
The result (\ref{8}) is plotted in fig.1a (upper branch)  %
and fig.1b (lower branch).  %
As long as the inequalities (\ref{A2}) are satisfied the squared  %
frequencies (\ref{8}) are both positive. At the borders of this region %
in parameter space the frequency $\omega_{b-}$ vanishes and $\omega_{b+}$  %
approaches $\Omega_1$ or $\Omega_2$ depending on which of the two borders has %
been reached. This can be seen  %
happening in figs.1a,b, where $\omega_{b+} = \Omega_2$ at the stability border  %
defined by the vanishing of $\omega_{b-}$ for the choice $\Omega_2^2=2\Omega_1^2$  %
made in these figures.

For axially symmetric trap potentials,   %
\begin{eqnarray} 
\label{9}%
U_i (%
\mbox{\boldmath $x$}%
) = \frac{m}{2}\left [  %
\Omega ^2_{\perp i} + \left (  %
\Omega ^2_{\parallel i} - \Omega ^2_{\perp i} %
\right )\cos ^2\theta  %
\right ] r^2 , %
\end{eqnarray}
with %
$\Omega_{\parallel 2}^2/\Omega_{\parallel 1}^2=\Omega_{\perp 2}^2/\Omega_{\perp 1}^2$ %
special solutions of Eqs. (\ref{7}) can still be  %
obtained, as in the single component case \cite{6}.  %
Solutions $\propto r^l Y_{lm}(\theta ,\phi)$ can be found for   %
\begin{eqnarray} 
\label{10}%
m=\pm l : \ \ \omega ^2_{+,-} = l \omega ^2_{\perp \, +,-}, %
\end{eqnarray}
where  %
\begin{eqnarray} 
\label{11}%
\omega ^2_{\perp \, +,-}= \left . \omega ^2_{b \, +,-}  %
\right | _{\Omega ^2_i \rightarrow \Omega ^2_{\perp \, i}}; %
\end{eqnarray}
and %
\begin{eqnarray} 
\label{12}%
m=\pm (l-1) : \ \ \omega ^2_{+,-} = (l-1) \omega ^2_{\perp \, +,-}+ %
\omega ^2_{\parallel \, +,-}, %
\end{eqnarray}
where  %
\begin{eqnarray} 
\label{13}%
\omega ^2_{\parallel \, +,-}= \left . \omega ^2_{b \, +,-}  %
\right | _{\Omega ^2_i \rightarrow \Omega ^2_{\parallel \, i}}. %
\end{eqnarray}
For the four coupled $m=0$ modes with $n=0$, $l=2$ and $n=1$, $l=0$  %
one obtains the four frequencies,   %
\begin{mathletters} %
\label{14}%
\begin{eqnarray} 
\omega ^2_{+,-, \, 1} = \omega ^2_{\perp \, +,-}\left (  %
2 + \frac{3}{2}\lambda _{+,-}^2 + \frac{1}{2} %
\sqrt{9 \lambda ^4_{+,-} - 16 \lambda _{+,-}^2 + 16} %
\right ), \nonumber \\%
\omega ^2_{+,-, \, 2} = \omega ^2_{\perp \, +,-}\left (  %
2 + \frac{3}{2}\lambda _{+,-}^2 - \frac{1}{2} %
\sqrt{9 \lambda ^4_{+,-} - 16 \lambda _{+,-}^2 + 16} %
\right ), %
\end{eqnarray}
\end{mathletters}
with $\lambda ^2_{+,-} = \omega ^2_{\parallel \, +,-}%
/\omega ^2_{\perp \, +,-}$.  %

Generalisations of our results for mixtures with more than two components can %
be easily  %
made. All that changes is the expressions for the prefactors $\omega ^2_b$ in %
the dispersion  %
law (\ref{0}). E.g. for condensates of ternary mixtures in isotropic %
concentric potentials  %
$U_i(%
\mbox{\boldmath $x$}%
)$ one would obtain three branches Eq. (\ref{0}),  %
whose squared frequencies $\omega ^2_{b\, 1,2,3}$ are the roots of a cubic equation and  %
depend on the ratios of all six scattering lengths  %
$a_{1},a_{2},a_{3},a_{12},a_{13},a_{23}$.  %

Mathematically, all explicit results (\ref{0}),  (\ref{10}),  (\ref{12}),  (\ref{14})  %
are the same as the results for the one-component condensate,  %
if there the oscillation frequency $\omega ^2_0$  %
in the trap is replaced by the two frequencies (\ref{8}). %
Physically, this difference  %
is of interest, because %
these frequencies are determined dynamically by the coupling between %
the two components in the binary phase.%

Measuring the frequency of the low lying excitations  %
of the binary condensate phase in coexistence with %
one or both single-component %
phases with basically the same experimental set-up  %
as employed in Refs. \cite{5,6} seems very feasible. %
Because of the coupling of the coexisting phases by particle number %
conservation the mode spectrum is predicted to depend in general %
on the geometry of the %
interphase boundaries. To check our prediction for %
the simple mode spectrum in the pure binary phase the following %
requirements have to be met: %
The ratios of the scattering lengths must be known %
and inequality (\ref{A2}) must happen to be satisfied %
for the atomic species under %
consideration. %
Then in principle a pure binary phase condensate can be produced in concentric %
spherically or axially symmetric traps assuming that the ratios of the particle %
numbers can be chosen appropriately. To obtain concentric traps, their %
potentials must be made %
sufficiently stiff to suppress the offset of their centers %
due to gravity. Satisfying all these conditions one should be able %
to observe that the mode spectrum reduces to the simple form (\ref{0}) with the %
scale factors (\ref{8}). %
%

\section*{Acknowledgements} %
We would like to thank Scott Parkins for assistance with the figures. %
This research ws supported by the Marsden Fund of the Royal Society of New Zealand %
and the New Zealand Lottery Grants Board.
One of us (R.G.) wishes to thank  %
the Deutsche %
Forschungsgemeinschaft for financial support through the  %
SFB 237 Unordnung und grosse Fluktuationen. %
 %
\begin{figure} %
\caption{The frequency scale factors 
$\omega ^2_{b+}/\Omega_1^2$ (figure 1.a) and $\omega ^2_{b-}/\Omega_1^2$
(figure 1.b) as a function  
of the ratio of the scattering lengths $a_{12}/a_1$, for different
values of $a_{2}/a_1$,  
for $\Omega ^2_2/\Omega ^2_1 = 2$. 
The chosen value of the ratio $\Omega ^2_2/\Omega ^2_1$ corresponds  
to the experiment \protect \cite{1}.} 
\label{Fig1}
\end{figure} %
\end{document}